\documentclass[conference]{IEEEtran}
\IEEEoverridecommandlockouts
\usepackage{cite}
\usepackage{amsmath,amssymb,amsfonts}
\usepackage{algorithmic}
\usepackage{graphicx}
\usepackage{textcomp}
\usepackage{xcolor}
\def\BibTeX{{\rm B\kern-.05em{\sc i\kern-.025em b}\kern-.08em
    T\kern-.1667em\lower.7ex\hbox{E}\kern-.125emX}}
\begin{document}

\title{Computational Perspective of the Fog Node}

\author{\IEEEauthorblockN{João Bachiega Jr, Breno Costa, and Aletéia P. F. Araujo}
		\IEEEauthorblockA{Department of Computer Science - University of Bras\'{i}lia (UnB) - Brasília - DF - Brazil \\
			Email: joao.bachiega.jr@gmail.com, brenogscosta@gmail.com, aleteia@unb.br}
	}

\maketitle

\begin{abstract}
Fog computing is a recent computational paradigm that was proposed to solve some weaknesses in cloud-based systems. For this reason, this technology has been extensively studied by several technology areas. It is still in a maturing stage, so there is no consensus in academia about its concepts and definitions, and each area adopts the ones that are convenient for each use case. This article proposes a definition and a classification relying on a computational perspective for the “fog node”, which is a fundamental element in a fog computing environment. In addition, the main challenges related to the fog node are also presented.
\end{abstract}


\section{Introduction}

Fog computing is a paradigm that enables provisioning of resources and services at the network edge, closer to the end devices. It is both complementary to, and an extension of, the traditional cloud-based model, addressing several issues of connected devices, such as high bandwidth, geographical dispersion, and low latency needs \cite{Yousefpour2019}.

The advantages of a model like fog computing drew the attention not only of the computing area, but also of several other ones, such as telecommunications and electrical engineering. In this perspective, the development of this computational paradigm, as well as the maturation of the concepts and definitions that surround it, are strongly linked to the different use cases of each study area.

Among these concepts is the term ``fog node'', which is a fundamental element in fog computing architecture. The concept of a fog node is wide and although some works have recently been published proposing taxonomies and definitions for it \cite{Naha2018}, \cite{hong2019resource}, there is no work, to the best of our knowledge,  that has contextualized the fog node in a computational perspective. 

Based on an analysis of some publications currently in the academy, this paper aims to contextualize the function of a computational resource in fog computing, as well as to present a definition and a classification for the fog node. So, Section \ref{sec:Fog} contextualizes fog computing, presenting the architecture's main characteristics. Section \ref{Sec:CompResource} takes an approach to the computational resource, presenting the desirable characteristics of a fog node and its forms of delivery, as well as presenting a definition to be adopted. Related works are presented and compared with our proposal in Section \ref{Sec:Related}. The challenges inherent to fog computing and fog nodes are presented in Section \ref{Sec:Desafios}. Finally, Section \ref{sec:Conclusion} brings the conclusions of this article and the future work that will be done next.

\section{Fog Computing} \label{sec:Fog}

Fog Computing is a distributed computing paradigm, integrated into the cloud, whose processing is done at the network edge. It provides computing resources for applications that cannot perform properly with the high latency provided by cloud-only environments \cite{Naha2018}.

Bonomi et al. \cite{Bonomi2012} presented the first definition of fog computing stating that it is a highly virtualized platform that provides computing, storage, and networking services among many computing data centers or end-devices. These components may or may not be at the edge of the network.

Several researchers have expanded and revised this initial definition of fog computing. Yi et al. \cite{Yi2015} consider fog computing as a scenario, composed of a high number of decentralized and heterogeneous devices, where they communicate and cooperate among themselves and with the network to perform data processing and storage without third-party interventions. Services, applications, or basic network functions that run in a sandboxed environment, can be supported by the data processing and storage.

For Dastjerdi et al. \cite{Dastjerdi2016}, fog computing is considered a distributed computing paradigm. In this paradigm,  the services provided by the cloud are essentially extended to the network edge. Fog computing addresses application requirements that need low latency with a huge and dense geographical distribution. Therefore, fog computing supports computing resources, different communication protocols, mobility, interface heterogeneity, integration with the cloud, and distributed data analytics.

For Naha et al. \cite{Naha2018}, fog computing is a distributed platform where the edge or end devices, that can be virtualized or not, will do the majority of processing. It resides in between the cloud and users and the cloud will do long-term storage and non-latency-dependent processing.

In the industry point of view, Cisco \cite{CiscoFog} defines that the fog extends the cloud to be closer to the things that produce and act on Internet of Things (IoT) data. These devices, called fog nodes, can be any device with computing, storage, and network connectivity and are deployed anywhere with a network connection. For IBM \cite{IBMFog}, the term ``fog computing'' and ``edge computing'' carry the same meaning. They both mean operation on network ends rather than hosting and working from a central cloud. They represent the scenario where  processes and resources are located at the edge of the cloud and do not establish any channel for cloud utilization.

The Open Fog Consortium \cite{OpenFog17}, which since January 2019 has merged with the Industrial Internet Consortium, states that fog computing is a decentralized computing infrastructure that considers the best place between the cloud and the data source to distribute computation, storage, and applications. It is both complementary to, and an extension of, traditional cloud-based models.

Finally, fog computing was defined by the National Institute of Standards and Technology (NIST) as a layered model facilitating the deployment of applications and services that are latency-aware and distributed. This model enables ubiquitous access to shared devices that are not perceptibly different from each other, although the extremes are quite distinct, from scalable computing resources. Therefore, fog computing provides, for the end-devices, local computing resources, network connectivity to centralized services and minimizes the request-response time, when needed.

In all definitions, it is noticeable that fog computing is tightly linked to the existence of a cloud, since fog can never replace the cloud completely as we still need it to handle big or complex data problems \cite{Gill2019}. Fog computing is suitable to be used when the cloud does not reach the time limit, bandwidth limitations, or latency requirements. 

\subsection{Architecture}

\begin{figure*}[ht]
	\centering
	\includegraphics[width=0.67\textwidth]{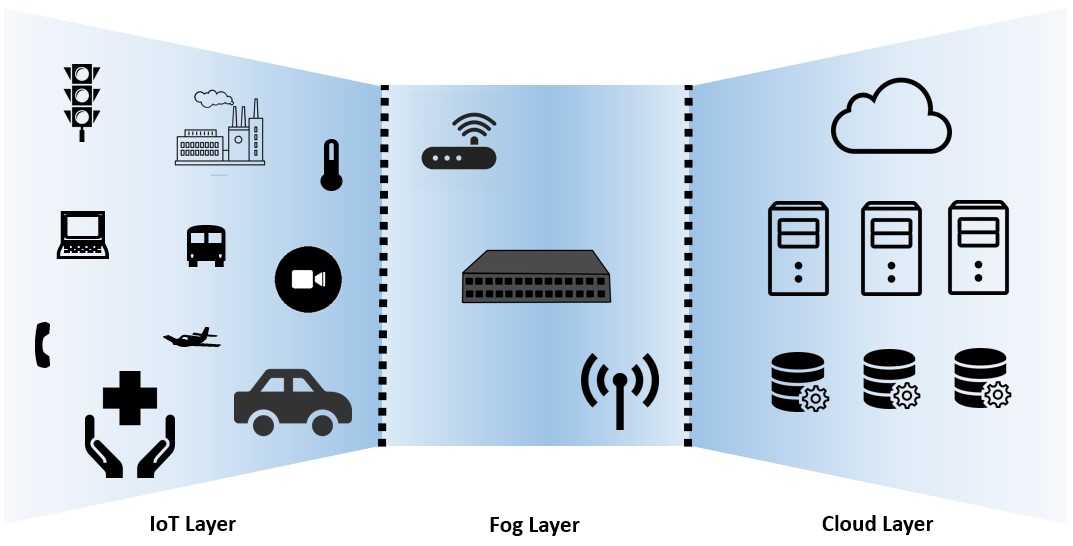}
	\caption{Horizontal fog computing architecture overview.}
	\label{fig:fogoverview}
\end{figure*}

The layered (or hierarchical) representation is the most widely used approach \cite{Naha2018}. In this context, a three-tiered architecture is the common representation of fog computing environment \cite{Mahmud16,Fan2017}. However, it is also possible to find proposals with four \cite{Tang2015}, five \cite{Naas2017} or even six \cite{Fan2017} layers. A comprehensive review of fog computing architectures can be found in \cite{Habibi2020}. So, a fog computing architecture composed of three layers will be used in this paper, namely: IoT Layer, Fog Layer and Cloud Layer, as presented in Figure \ref{fig:fogoverview}. 

Unlike other proposals that use hierarchical architectures, this work applies a horizontal architecture. The intention is to show that both the IoT Layer and the Cloud Layer, which are respectively on the left and right edges of Figure \ref{fig:fogoverview}, have a number of devices and resources that tend to the infinite \cite{Yousefpour2019}, while the internal Fog layer has a more limited number of devices and computational resources.

The IoT layer represents all IoT devices connected at the edge of the network. It is in this layer that end-users request services that will be processed in the Fog and Cloud layers.

The Fog layer acts as a link between the IoT and Cloud layers to provide the necessary functionalities for data processing applications, such as filtering and aggregation, before transferring the data to the cloud \cite{AlDoghman2016}. This layer is composed of nodes, also called \textit{Fog Nodes}, and comprises ``smart'' devices capable of processing and storing data, in addition to routing and forwarding data packets to the Cloud Layer. A detailed explanation of fog nodes is given in Section \ref{Sec:CompResource}.

Finally, the Cloud layer has a more robust computational resources to process all requests made by the IoT layer, which have not been fully answered by the Fog layer. Therefore, the existence of the Cloud layer is fundamental in a fog computing environment \cite{AlDoghman2016}.

\subsection{Essential Characteristics}

Regarding the characteristics of fog computing, although some publications are divergent \cite{Bonomi2012}, \cite{Mukherjee2018} and \cite{Hu2017}, the ones most used by academics are those indicated by the NIST \cite{fognist}, described bellow:

\begin{itemize}
	\item Low Latency: as fog computing devices are often located at the edge of the network, the analysis and response to data generated by these devices is much faster than from a cloud service;
	\item Geographic Distribution: in contrast to cloud computing, services and applications driven by fog computing require widely geo-distributed deployments;
	\item Heterogeneity: supports the collection and processing of data originating from different sources and acquired through various types of network communication resources;
	\item Interoperability and Federation: components must be able to interoperate and services must be federated between domains;
	\item Real-Time Interactions: fog computing applications involve real-time interactions, rather than batch processing;
	\item Scalability: supports scalability of computational resources, changes in workloads and variations in network and device conditions.
\end{itemize}

Considering these characteristics, fog computing is suitable to use when cloud computing does not provide the latency or runtime requirements required by applications \cite{Gill2019}. 

\subsection{Related Paradigms}

Beyond Fog Computing, there are other related paradigms and technologies with similar purposes, but with different architectures and characteristics. As with fog computing, for these other paradigms there is still no mature definition by the academy, which generates some confusion regarding how much one paradigm makes an intersection with the others that also has similar characteristics.

Fog computing is often erroneously confused with edge computing, but there are key differences between the two concepts. Fog computing runs applications in a multi-layer architecture, while edge computing runs specific applications in a fixed logical location and provides a direct transmission service. The edge computing tends to be limited to a few numbers of peripheral devices \cite{fognist}, whereas fog computing, in general, has a bigger number of peripheral devices and is hierarchical.

In addition to edge computing, there are in literature, some variations and other proposals for paradigms and technologies that aim to perform tasks in an environment closer to the device. Mobile Edge Computing, Mobile Cloud Computing, Mobile Ad Hoc Cloud Computing, Mist Computing, Cloudlet Computing, and Dew Computing are examples of them. 
In any case, supported by the definition of \cite{stojmenovic2014fog}, ``some other concepts, not declared as `fog computing’, might fall under the same `umbrella’ ''. In the academy it is possible to find publications focused on the presentation and comparison of fog computing with these other paradigms and computational technologies, such as \cite{Yousefpour2019}, \cite{Naha2018} and \cite{Mukherjee2018}.

\section{Computational Resources in Fog Computing} \label{Sec:CompResource}

By the computational perspective, a resource is any physical or virtual component \cite{Manvi2014}, e.g. CPU, memory, data, network devices, operating systems, virtualization systems, etc.

Specifically in fog computing, it is common to use the term ``\textit{Fog Node}'' \cite{fognist} to describe an element that composes the architecture, located in the Fog layer. Due to the lack of a widely adopted definition for this term, the concept may vary according to the use case and the application area to which the fog node is applied. Examples for this are the proposals for areas of energy, health, telecommunications, smart buildings, vehicular networks, among others \cite{Yousefpour2019}.

Unlike other computational paradigms, such as cloud or grid computing -- in which computational resources have high processing and storage capacity, are more linear in terms of architecture and compatibility, and still be located in data centers installed in specific points of the world -- the devices that compose fog computing, in general, have less computational capacity, are more heterogeneous \cite{Yousefpour2019} and are widely geographically distributed \cite{hong2019resource}.

So far, to the best of our knowledge, there is no publication or reference architecture about the hardware parameters to classify some device as a fog node. Therefore, considering that the requirements of fog computing devices are closer to IoT Layer devices than to the Cloud layer, we searched the references on IoT for the necessary information to support this definition. A comprehensive analysis about IoT reference architectures can be found in \cite{weyrich2015reference}.

\begin{figure}[h!]
	\centering
	\includegraphics[width=0.35\textwidth]{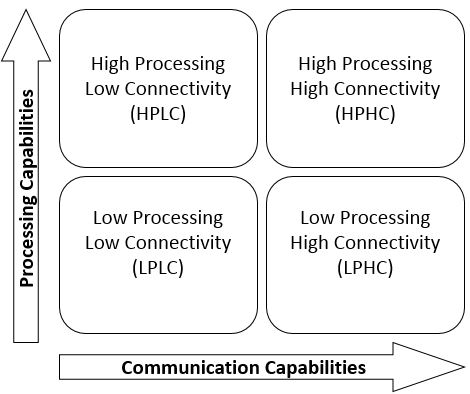}
	\caption{Types of devices correlating the processing and communication capabilities \cite{ITU4460}.}
	\label{Fig:FiguraITU}
\end{figure}

Among these publications, the International Telecommunication Union (ITU) architectural reference models of devices for IoT applications \cite{ITU4460} states that ''a device is a piece of equipment with the mandatory capabilities of communication and optional capabilities of sensing, actuation, data capture, data storage, and data processing". Internet Engineering Task Force (IETF)'s RFC 7228\footnote{https://tools.ietf.org/html/rfc7228} classified devices into four types, correlating the processing and communication capabilities, as show in Figure \ref{Fig:FiguraITU}.

It is important to note that ITU's recommendation \cite{ITU4460} states that as the connectivity capability also depends on processing capability, the combination of high processing and low connectivity (HPLC area on Figure \ref{Fig:FiguraITU}) is not usual. Thus, this type of device will not be considered in our definition.

A brief description of the remaining classifications is presented below:

\begin{itemize}
    \item \textbf{Low Processing and Low Connectivity (LPLC) device:} this type of device has not sufficient processing capabilities to make decisions or run complex algorithms and also does not directly connect to a communication network. These devices must rely on other network element. Examples includes sensor nodes, like Waspmote and Meshlium \cite{perera2017fog};
	\item \textbf{Low Processing and High Connectivity (LPHC) device:} although these devices also have little processing power, they can directly communicate with applications or cloud services through the Internet. Routers, gateways, set-top boxes and access points are examples of this type of device \cite{li2018edge};
	\item \textbf{High Processing and High Connectivity (HPHC) device:} in addition to having the ability to execute more complex applications and algorithms and having a direct connection to the internet, they can also directly coordinate other devices. Some examples includes sink nodes (e.g.Raspberry Pi), low-end computational devices (e.g. smartphones) and high-end computational devices (e.g. personal computers) \cite{perera2017fog}.
\end{itemize}

\begin{figure*}[h!]
	\centering
	\includegraphics[width=0.68\textwidth]{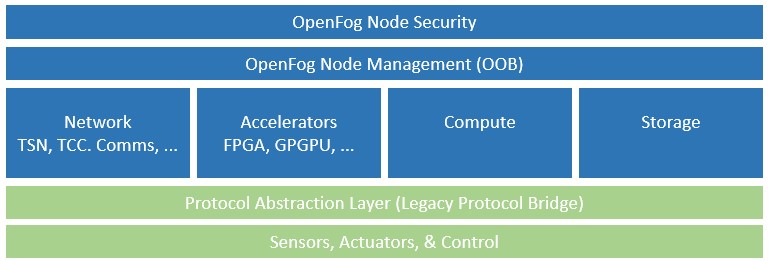}
	\caption{Fog node structure overview proposed by the OpenFog Consortium \cite{OpenFog17}.}
	\label{Fig:FogNodeOpenFog}
\end{figure*}

In addition to the processing and communication capabilities, it is necessary to indicate what are the expected functional requirements for a fog node. In this sense, there are some definitions already published. Considering the reference architecture proposed by the OpenFog Consortium \cite{OpenFog17}, a fog node is composed of 8 aspects, as shown in Figure \ref{Fig:FogNodeOpenFog}.

The desirable characteristics of a fog node proposed by \cite{OpenFog17} and presented in Figure \ref{Fig:FogNodeOpenFog} are: 

\begin{itemize}
	\item \textbf{Security:} node security is essential to the overall security of the system. This includes protection for interfaces, computing, software, etc;
	\item \textbf{Management:} a node must support management interfaces to allow top-level system agents to see and control the lowest-level node. The same management protocol can be used on many different physical interfaces;
	\item \textbf{Network:} must provide the scalability, availability and flexibility required by Quality of Service (QoS) in addition to prioritizing critical or latency-sensitive data. Depending on the deployment scenario, the fog nodes are likely to be the network element itself, such as an access point, a gateway or a router;
	\item \textbf{Accelerators:} some fog nodes can be allocated for enhanced analysis, that cannot be served by a conventional CPU. In these cases, the accelerator modules will be configured next to the processor modules (or integrated with them) to provide additional computational performance;
	\item \textbf{Computing:} a node must have general-purpose computing resources to allow a higher level of interoperability;
	\item \textbf{Storage:} many types of storage may be needed on fog nodes. Thus, storage layers normally seen in datacenters, such as RAM Arrays, Solid State Drives and Fixed Spinning Disks must be supported in a fog node;
	\item \textbf{Sensors and Actuators:} these devices based on hardware or software are considered to be the lowest level elements in the IoT layer. There may be several hundred or more of these devices associated with a single fog node. Some of them are ``dumb'' devices, without any significant processing capacity, while others may have some basic computing functions;
	\item \textbf{Abstraction Protocol:} many of the sensors and actuators on the market today are not able to interact directly with a fog node. The abstraction protocol layer makes it possible to logically place these elements under the supervision of a fog node so that their data can be used for analysis and higher-level system functions.
\end{itemize}

The aspects presented by the Open Fog Consortium \cite{OpenFog17} try to encompass many kinds of devices that can be part of a fog environment, but they do not have a focus exclusively on the computing area. Although some items are highly recommended, such as network, computing and the abstraction protocol, other items are not found in most fog computing devices, such as accelerators.

In addition, in the document proposed by NIST \cite{fognist}, the following attributes are considered for fog nodes:

\begin{itemize}
	\item \textbf{Autonomy:} fog nodes can operate independently, making local decisions at the node level or at the cluster of nodes;
	\item \textbf{Heterogeneity:} fog nodes have different formats and can be deployed in a wide variety of environments;
	\item \textbf{Hierarchical grouping}: fog nodes support hierarchical structures, with different layers providing different subsets of service functions;
	\item \textbf{Manageability:} fog nodes are managed and orchestrated by systems that can automatically execute routines;
	\item \textbf{Programmability:} fog nodes are programmable at various levels, by various stakeholders - such as network operators, domain specialists, equipment providers or end-users.
\end{itemize}



\begin{figure}[h!]
	\centering
	\includegraphics[width=0.41\textwidth]{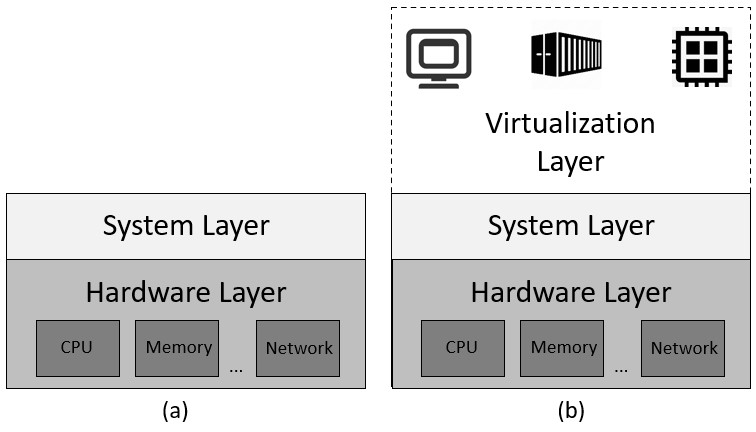}
	\caption{Overview of a basic fog node (a) and a fog node with virtualization (b). Adapted from \cite{marin2017we}.}
	\label{Fig:FogNode}
\end{figure}

Taking into account the characteristics, attributes and functionalities presented so far, it is possible to state that, 
by a computational perspective, a fog node is composed of a Hardware layer - where the physical resources (e.g. CPU, memory, network interface, among others) are located - and also a System layer, necessary for the abstraction of hardware and the execution of applications \cite{coulouris2000distributed}, as shown in Figure \ref{Fig:FogNode} (a).

Also, based on the list of attributes and functional requirements presented, only devices classified as LPHC and HPHC are adherent to fog computing, since the devices classified as LPLC are unable to meet the indicated requirements.

In a computational perspective, this is confirmed because a desirable feature of a fog node is the virtualization capability \cite{mann2020notions} and it cannot be implemented by LPLC devices. Virtualization can be defined as ``a software abstraction with the appearance of a computer system hardware'' \cite{goldberg1974survey}. In addition, hardware-based virtualization mechanisms are available on almost all processor hardware that would be used to implement fog platforms \cite{OpenFog17}. In this logic, for a better use of the hardware resources available in the ``Basic Fog Node'' (Figure \ref{Fig:FogNode} (a)), the existence of a Virtualization layer is desirable \cite{mann2020notions}, as shown on Figure \ref{Fig:FogNode} (b).

Thereat, all the four attributes defined by NIST \cite{fognist} - autonomy, heterogeneity, manageability and programmability - can be reached in a simple way, once the operationalization is done by an abstract management layer.

The most common way of delivering virtual resources is through virtual machines \cite{goldberg1974survey}, often used in cloud computing environments, since it provides the ability to split up resources of a physical machine and reallocate them in one or more virtual machines, which can be used in an infinite number of tasks. 

More recently, the virtualization method has migrated to the use of containers since they offer an isolation mechanism more adherent to a fog computing environment. Containers perform virtualization at the operating system layer and no longer at the hardware layer, as occurs with virtual machines \cite{hong2019resource}. Container-based virtualization acts like sand-box operating systems, with a host operating system running on the bottom layer and  sharing its kernel in read-only mode. A great advantage of containers is that they can run the platform independently, due to the fact that they share the host kernel, favoring, for example, the migration of a server.

Another form of virtualization that has  emerged in recent publications on fog computing \cite{cozzolino2020ecco} is called \textit{unikernel} \cite{madhavapeddy2014unikernels}. It is a type of application-level virtualization that is even leaner than containers. An unikernel is a specialized operating system, written in high-level code, that is compiled from the application code. The drivers and operating system functions that are included in the compilation are only the ones needed to run the application, making them extremely lightweight and very fast as there are no context switches \cite{madhavapeddy2014unikernels}. 

\begin{figure*}[h!]
	\centering
	\includegraphics[width=0.67\textwidth]{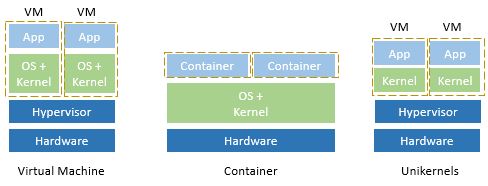}
	\caption{Architecture representation of virtual machine, container and unikernel. Adapted from \cite{Mouradian2018}.}
	\label{Fig:VMxContainerxUnikernel}
\end{figure*}

A diagram comparing the three ways of virtualization (virtual machine, container and \textit{unikernel}) is shown in Figure \ref{Fig:VMxContainerxUnikernel}.

Finally, considering the scope and diversity of devices that can be part of a fog computing environment as presented in this section, and because there is still no standardization adopted by the academy, the following definition of a fog node is proposed with the objective of limiting the scope and of being more adherent to the objectives of a computational perspective: ``\textit{A fog node is any hardware device in a fog computing environment that has system and hardware resources added to high communication capability.}''.

However, from a computational perspective, the capacity for virtualization is considered important to make possible the operationalization of the computational resource, allowing the execution of applications to occur even in the Fog layer.

\begin{figure}[h!]
	\centering
	\includegraphics[width=0.33\textwidth]{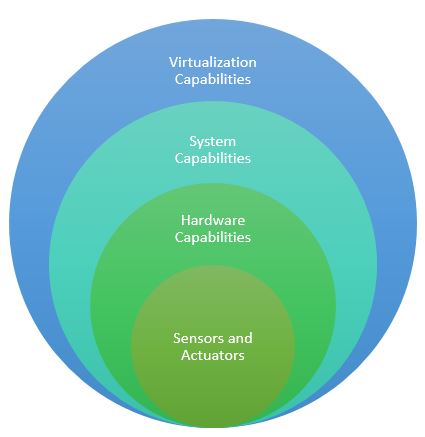}
	\caption{Fog node classification by a computational perspective.}
	\label{Fig:Classificacao}
\end{figure}

Based on our fog node definition, considering a minimal approach from a computational perspective, Figure \ref{Fig:Classificacao} shows the  classification proposed and in this onion diagram (Figure \ref{Fig:Classificacao}), it is possible to notice that the nucleus is composed of \textit{sensors and actuators} that are part of the IoT Layer (as shown in  Figure \ref{fig:fogoverview}). As the onion layers expand, they aggregate \textit{hardware capabilities}, that must be at least CPU, memory and network and can also include accelerators like GPU; \textit{system capabilities} like operation system, and, finally,  \textit{virtualization capabilities}, where is possible some type of virtualization, such as virtual machine, container or unikernel.

With this in mind, we have that the more layers the fog node covers, the more favorable it will be for applications and computing services, so it would be possible to meet several other requirements, such as security, management and programmability.

\section{Related Work} \label{Sec:Related}

Fog computing is a research subject on the rise in both academia and industry. In recent years, several works have been published about this computational paradigm. Many of these works are submitted in the form of literature reviews and surveys, such as \cite{Yousefpour2019} and \cite{Mukherjee2018}.

Some of these works have addressed concepts related to computational resources in fog computing. In \cite{Mahmud16}, which was one of the first published papers of this type on fog computing, the authors describe five types of fog nodes: servers, networking devices, cloudlets, base stations and vehicles. The authors \cite{Mahmud16} conceptualize fog nodes as ``computational nodes with heterogeneous architecture and configurations that are capable to provide infrastructure for Fog computing at the edge of the network''. But it is important to note that in these authors' point of view a ``Fog computing environment is composed of traditional networking components e.g. routers, switches, set top boxes, proxy servers, base stations'' and these components are provided with diverse computing, storage, networking, etc. capabilities and can support service-applications execution.

The paper \cite{Naha2018} presents a taxonomy that classifies the devices into three categories: IoT devices, processing devices and gateway devices. Thus, it is possible to make a relationship between the classification proposed by Naha et al. and the three layers of the fog computing architecture presented in Figure \ref{fig:fogoverview}. Considering this relationship, IoT devices are allocated in the IoT layer, processing devices are allocated in the Cloud layer and gateway devices are allocated in the Fog layer. The paper \cite{Naha2018} also presented infrastructure (processing, storage, network and memory) and network (connection and mobility) requirements, but no fog node definition was proposed.

In the work \cite{marin2017we}, fog nodes are classified as ``smart'' -- for those devices that have some computational processing and storage capacity -- and ``dumb'', for those devices that have limited computational capacity, such as sensors and actuators. The following fog node definition is presented by these authors: ``Fog nodes are distributed fog computing entities enabling the deployment of fog services and formed by at least one or more physical devices with processing and sensing capabilities.'' The authors complement their definition indicating that ``all physical devices of a fog node are connected by different network technologies, aggregated and abstracted to be viewed as one single logical entity, i.e., the fog node, able to seamlessly execute distributed services, as they were on a single device''.

A representation model for physical resources in a fog environment is presented in \cite{abouaomar2019resources}, focusing on processing, storage, memory and networking capabilities. The authors also consider that fog computing generally relies on virtualization because of the capacity of virtualization to represent heterogeneous systems, but in their proposal, they just considered virtual machines and containers. No fog node definition was presented by these authors.

Finally, within a computational perspective, in the paper \cite{hong2019resource}, resources are classified more broadly into hardware and software. Hardware resources are made up of computing devices and network devices, while software resources are made up of virtualization systems and virtualized networks. Hong and Varghese \cite{hong2019resource} stated that devices in the fog must have not only physical resources available, but also virtual resources that allow them to operate and manage the hardware resources. No fog node definition was presented by these authors.

Table \ref{Tab:RelatedWork} compares the related work presented in this section with the goals of this paper. This study is the only one that provides information about computational resources in a fog environment, that is, the fog node, in a computational perspective, exploring not only the physical attributes but also virtual ones, including unikernel, that is a lightweight virtualization mode strongly adherent to fog computing purposes \cite{cozzolino2020ecco}.

\begin{table*}[!ht]
	\footnotesize
	\centering
	\caption{Related work comparison.}
\begin{tabular}{|c|c|c|c|c|c|c|}
\hline
\textbf{Paper}                  & \multicolumn{1}{l|}{\textbf{\begin{tabular}[c]{@{}l@{}}Fog node\\ definition\end{tabular}}} & \textbf{\begin{tabular}[c]{@{}c@{}}Computational\\ Perspective\end{tabular}} & \textbf{Virtualization} & \textbf{\begin{tabular}[c]{@{}c@{}}Virtual\\ Machine\end{tabular}} & \textbf{Container} & \textbf{Unikernel} \\ \hline
\cite{Mahmud16}               & \checkmark                                                                                           & \checkmark                                                                         &                         &                                                                    &                    &                    \\ \hline
\cite{Naha2018}               &                                                                                             & \checkmark                                                                         &                         &                                                                    &                    &                    \\ \hline
\cite{marin2017we}            & \checkmark                                                                                           &                                                                           & \checkmark                       & \checkmark                                                                  & \checkmark                  &                    \\ \hline
\cite{abouaomar2019resources} &                                                                                             & \checkmark                                                                         & \checkmark                       & \checkmark                                                                  & \checkmark                  &                    \\ \hline
\cite{hong2019resource}       &                                                                                             & \checkmark                                                                         & \checkmark                       & \checkmark                                                                  &                    &                    \\ \hline
\textbf{This Work}               & \checkmark                                                                                           & \checkmark                                                                         & \checkmark                       & \checkmark                                                                  & \checkmark                  & \checkmark                  \\ \hline
\end{tabular}
\label{Tab:RelatedWork}
\end{table*}

\section{Challenges} \label{Sec:Desafios}

Although interest in fog computing is growing nowadays, some old challenges related to distributed systems still have to be addressed. These challenges are heterogeneity, openness, security, scalability, failure handling, concurrency, transparency, and QoS \cite{coulouris2000distributed}. There still several challenges that still need work to reach the state of art.

Up until now there has been no standard fog computing architecture \cite{Habibi2020}, although some organizations have been trying to achieve standardization, such as OpenFog Consortium \cite{OpenFog17} or NIST \cite{fognist}. Due to this lack of standardization, proposed works using fog computing are not applicable for all use cases and consequently, it is not possible to measure and improve metrics, like Service Level Agreement (SLA) or QoS. Furthermore, there is no standardization for SLA or QoS applicable for fog computing systems, since the current ones were defined for cloud services or network infrastructure. Fault tolerance aspects also need to be considered \cite{Bilal2018}.

The absence of a standard programming platform is also a point of concern \cite{Hu2017} since data-processing frameworks are based on static configurations rather than dynamic ones. Dynamic configurations are necessary for fog environments that require the ability to effectively add and remove nodes.

Integration with brand new technologies is another relevant open issue in fog computing environments. Among them are the use of Artificial Intelligence (AI) and Blockchain \cite{Gill2019}. These technologies could be used in relevant tasks, such as predicting the requirements in advance for geographically dispersed resources and also proper task scheduling on heterogeneous fog environments \cite{Gill2019}. The use of blockchain concepts could improve security in fog environments \cite{kim2019novel} since it could be executed on resource constrained devices and this is a fog computing characteristic.

Security is considered to be a relevant aspect that is still a challenge for fog computing. The use of distributed and mobile devices  is a topic that deserves a lot of attention. Security and privacy issues, like man in the middle, authentication, distributed denial of service, access control \cite{Hu2017}, network security, and location verification are necessary to ensure data security, thereby expanding the number of use cases for fog computing.

Energy management is another topic that has been widely discussed \cite{li2019energy}. The challenge is to develop network protocols and architectures that can cope with the fog features, considering the high number of distributed nodes, and decreasing energy consumption. The use of fog computing can also reduce energy consumption in the cloud by lowering the volume of data processing in cloud data centers.

The lack of fog providers and billing models are also challenges that must be addressed \cite{Mahmud16}. Fog computing services may be provided by the Internet, Telecommunication, and Cloud service providers. Pricing and billing remain a challenge in terms of sustaining a commercial ecosystem of value-added services, as the business model is still not clear \cite{Yi2015}. Moreover, there are possible billing models such as consumption-based, where users are billed per usage, or subscription-based, where users pay a fixed monthly price and can use the fog on a network-wide basis \cite{bittencourt2015towards}.

Finally, a relevant aspect concerns the device's resources availability. Not necessarily a device with the required processing and communication characteristics has all of its resources available to run services at a specific moment. To keep availability information updated dynamically it is necessary to deploy a fog orchestration system. An orchestration system aims to monitor and manage the computational resources, dealing with heterogeneity, device mobility and connection uncertainty, providing services to the customers while achieving SLA and QoS requirements. Such orchestration system for fog environments is still in its infancy \cite{viejo2019secure}, although some papers have been published about it in recent years  \cite{donassolo2019fog}.

\section{Conclusions} \label{sec:Conclusion}

The distance of cloud computing data centers from the end user’s devices motivated the development of computational technologies that address this issue. Fog computing has proved to be one of the most robust and promising technology in use by several areas, like Computing and Telecommunications.

However, considering that this computational paradigm is still in its infancy, each area creates theories and definitions based on their own use cases, generating a mixture of concepts. Among these is the term ``fog node'', which is an essential element in a fog computing environment. In this sense, this work presented a contextualization about the computational resources in fog computing, highlighting the essential characteristics and, thus, a definition for ``fog node'' from a computational perspective was introduced. Works with the same objective, but which sought to contextualize and create definitions for other terms, such as the term \textit{architecture} are possible to find in \cite{mann2020notions} and also in \cite{Habibi2020}.

Among the characteristics presented, the need for a virtualization layer deserves to be highlighted, since it is fundamental to the management of computational resources and their best use. The virtualization models currently in use in fog computing are virtual machine, container and unikernel. Although the characteristics and requirements of each use case are decisive for defining the best form of virtualization to be adopted, in general, it is possible to state that the container and unikernel models are lighter and, therefore, more adherent to this computational model, which is basically composed of resources with low computational power.

As future work, the authors indicate extensive testing, such as provisioning time, migration, application execution time in a fog computing environment with the various virtualization models - virtual machines, container and unikernel. Although there are similar works in cloud computing  \cite{plauth2017performance}, there is not, to the best of our knowledge, some work that has carried out this analysis, addressing these three virtualization models specifically for fog computing.

\bibliographystyle{IEEEtran}
\bibliography{IPDPS21}

\end{document}